\documentclass{aip-cp}

\usepackage[numbers,elide]{natbib}
\usepackage{fancybox}
\usepackage{shortvrb}
\MakeShortVerb\|
\usepackage{amssymb}
\usepackage{graphicx}
\usepackage[caption=false]{subfig}
\graphicspath{{./graphics/}}
\usepackage{rotating}
\usepackage{listings}
\usepackage[usenames]{color}
\definecolor{codecolor}{gray}{.9}
\definecolor{rlcolor}{cmyk}{0,1,0,0}


\begin{document}

% Title portion
\title{Central Exclusive Production in the STAR Experiment\\at RHIC}

\author[aff1]{Rafal Sikora}
% \eaddress[url]{http://home.agh.edu.pl/$\sim$rsikora}
\eaddress{E-mail:~rafal.sikora@fis.agh.edu.pl}

\affil[aff1]{AGH University of Science and Technology, 30 Mickiewicza Ave, PL 30-059 Krakow, Poland}
% \affil[aff2]{Additional affiliations should be indicated by superscript numbers 2, 3, etc. as shown above.}
% \affil[aff3]{You would list an author's second affiliation here.}
% \corresp[cor1]{E-mail: rafal.sikora@fis.agh.edu.pl}

\maketitle

\begin{abstract}~The STAR experiment at the Relativistic Heavy Ion Collider (RHIC) performs studies of diffractive processes with the focus on the exclusive production of particles in central range of rapidity. In 2015 STAR collected 18~pb$^{-1}$ of data in polarized proton+proton collisions at $\sqrt{s}$~=200~GeV to measure Central Exclusive Production (CEP) process $pp\to pXp$ through Double Pomeron Exchange (DPE) mechanism. The intact protons moving inside the RHIC beampipe after the collision were measured in silicon strip detectors (SSD), which were placed in the Roman Pot vessels. This enables full control over interaction kinematics and verification of the exclusivity of the reaction by measuring the total (missing) transverse momenta of all final state particles: the central diffractive system in the Time Projection Chamber (TPC) and the forward protons in the Roman Pots. With the use of ionization energy loss in the TPC, 
d$E$/d$x$, it was possible to discriminate various production channels in $pp\to pXp$ reaction. This paper presents results on exclusive production of two charged particles ($\pi^{+}\pi^{-}$ and $K^{+}K^{-}$) in mid-rapidity region, $-1<\eta<1$, with small squared four-momentum transfer of forward protons, $0.03<-t<0.3~(\textrm{GeV}/c)^{2}$, obtained using 2.5\% of full statistics.\end{abstract}
% \linespread{0.95}

\section{INTRODUCTION}
% In high-energy physics a process is recognized as exclusive when a single interaction leads to production of a set of particles, of which all are measured in the detector. It is an interesting class of particle interactions, especially from the point of view of the theorists who have to account for all degrees of freedom involved in the process in order to thoroughly describe the exclusive production. Therefore exclusive processes are considered as an important tool to probe the structure of particles participating in the interaction.

A process is defined as exclusive when all particles in the final state, after the interaction, are measured and identified. Because of this, exclusive processes are of interest to the theorists. Those processes are also an important tool to probe the structure of particles participating in the interaction.

The STAR experiment~\cite{STAR} at RHIC - machine capable of colliding beams of heavy ions and polarized protons, conducts measurements of exclusive production in various colliding systems. In gold-gold interactions Ultra-Peripheral Collisions (UPCs) are studied~\cite{Klein2}. In these interactions, nuclei do not overlap, but act as sources of high flux of photons, which can produce lepton pairs ($e^{+}e^{-}$, $\mu^{+}\mu^{-}$) through $\gamma\gamma$ interaction. Also, in the UPCs photonuclear interactions between a photon from one nucleus and a Pomeron from another can lead to a vector meson production like $\rho^{0}$ or $J/\psi$. The last subprocess belongs to a~group of diffractive interactions, which are also studied in proton+proton collisions with an emphasis on the DPE, a fusion of two Pomerons in the Regge framework (see Ref.~\cite{Barone} for a~review).

The first evidence of the DPE was found at the Intersecting Storage Rings~\cite{EvidenceISR} at~CERN and since then it has been studied in numerous experiments. Large interest in this process is connected with the Pomeron picture in perturbative Quantum Chromodynamics (pQCD), in which at leading order Pomeron is represented by a pair of gluons (a color singlet). This gluon-rich environment of the reaction is commonly considered as suitable for production of the gluon bound states (glueballs), whose existence has not been confirmed unambiguously yet and which would be a strong argument for the validity of the QCD theory. Another important outcome of the DPE measurement is determination of the absorptive corrections, the probability that the rapidity gap(s) that charachterises diffractive interaction would be filled with particles produced in initial-~or final-state interactions. Last but not least, quantities defining the dynamics of the Pomeron exchange, i.e. the slope of the Mandelstam $t$ 
distribution or relative azimuthal angle between the scattered protons, could significantly contribute to extending knowledge of the diffraction in high-energy physics.

To measure CEP one needs the experiment with the ability of detecting particles produced in central region of rapidity and measuring protons scattered in the forward direction. Having forward proton detectors allows both efficient triggering and event selection. Most importantly, it allows full reconstruction of the kinematics of the final state. Consequently the partial wave analysis (PWA) of the state reconstructed at mid-rapidity can be performed with much reduced background. The STAR detector has all necessary components to widen our understanding of the exclusive diffraction.

\section{EXPERIMENTAL SETUP}

\begin{figure}[b]
\centering
\includegraphics[width=1.01\linewidth]{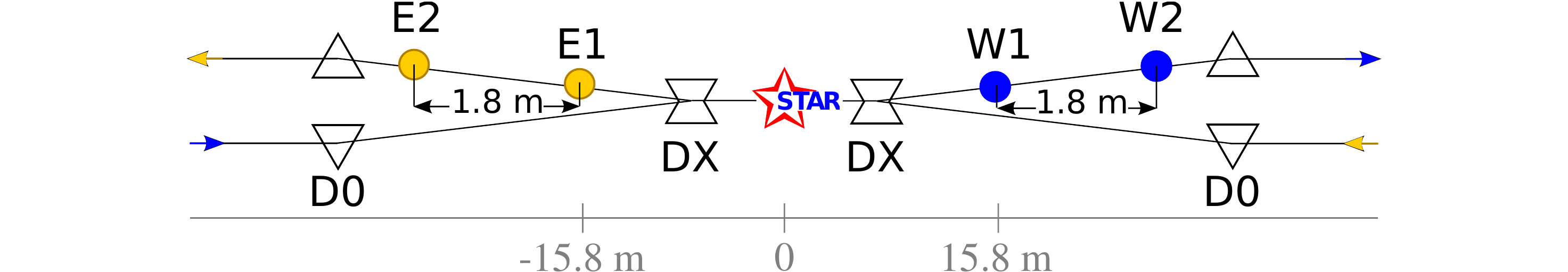}
\caption{Layout of the Roman Pot Phase II* subsystem (not to scale). On both sides of the central detector two Roman Pot stations are installed (E1 and E2 on the east, W1 and W2 on the west) between DX and D0 dipole magnets.}\label{Fig:rpScheme}
\end{figure}

The heart of the STAR detector is the TPC, a~4.2~m long cylinder with a diameter of 4~m, used for tracking charged particles with minimum transverse momentum of $100~\textrm{MeV}/c$ in pseudorapidity range $-1.2\lesssim \eta \lesssim1.2$.

The TPC is surrounded by the Time-Of-Flight (TOF) detector, a system of adjacent Multi-gap Resistive Plate Chambers. In addition to being precise timing detectors they are used to measure event multiplicity at the trigger level and discriminate TPC tracks arised in preceding bunch crossings from the in-time tracks. They also allow particle identification using timing information merged with momentum and path length reconstructed in TPC to determine the particle mass.

In the endcap part of the STAR detector there are forward scintillators installed, the Beam-Beam Counters (BBC). Extending between $2.1<\arrowvert\eta\arrowvert<3.3$ (large BBC tiles) and $3.3<\arrowvert\eta\arrowvert<5.0$ (small BBC tiles), they are mainly used to determine the rate of inealastic interactions and thus measure the instantaneous luminosity delivered by the machine.

In addition to above there are other important components, e.g. Zero-Degree Calorimeters (ZDCs) which are responsible for measuring neutral remnants leaving the collision at near-beam angle. However, attention should be put on the very forward proton detectors, which are a key subsystem for the diffractive measurements in STAR.

In 2015 the experiment was upgraded with the system of silicon strip detectors mounted in the Roman Pot vessels, which were originally used by the PP2PP experiment~\cite{PP2PP} designed to measure elastic proton-proton scattering at RHIC. The vessels separate the vacuum inside the beampipe from the air in their interior where SSDs are installed, and allow close approch of SSDs to the beamline to measure small scattering angles. Geometrical acceptance of the Roman Pot detectors limited values of squared four-momentum transfer from $0.03$ to $0.3~(\textrm{GeV}/c)^{2}$. With current detector arrangement, named the Roman Pot Phase II* and shown in Figure~\ref{Fig:rpScheme}, it allows to collect data with those forward detectors during regular high-luminosity runs and enables gathering large samples of data dedicated to diffractive processes.

As presented in Figure~\ref{Fig:rpScheme} the Roman Pot Phase II* setup consists of detectors located in two stations on each side of the interaction point (IP) in a distance of 15.8~m and 17.6~m from the IP. Each station has two Roman Pots positioned vertically, one above and the other below the beamline. Detectors are situated downstream the DX dipole magnets responsible for head-on targeting of the incoming beams and bending outgoing beams back into the accelerator pipeline. The constant and uniform magnetic field of the DX magnet works as a~spectrometer and thus knowledge on the track angle and position in the detector allows complete reconstruction of the proton momentum, including the fractional momentum loss $\xi$.

Single Roman Pot houses a package of 4 silicon strip detector planes - one pair of SSDs with vertical and one with horizontal orientation of the strips, and hence measurement of the position of a proton hit is possible in both transverse spatial coordinates, $x$ and $y$. The pitch (distance between neighbouring strips) in a single detector is 100~$\mu$m, therefore intrinsic spatial resolution is at the level of 30~$\mu$m. In addition to the silicon detectors, the package contains plastic scintillator that covers whole active area of the silicon, attached at the back. Two lightguides are glued at the top edge of scintillator which direct the light generated when ionizing particle passes through it to the photomultiplier tubes connected at the very end of each. This counter is used to trigger on forward protons and also provides the timing information that is used at the later stage of data analysis.\vspace*{-10pt}

\section{EVENT SELECTION AND ANALYSIS RESULTS}

Results described in this paper are based on 2.5\% sub-sample of data collected in year 2015 during 11-week period of $pp$ collisions at the center-of-mass energy $\sqrt{s}=200$~GeV. The STAR experiment collected 18~pb$^{-1}$ of data dedicated for the central diffraction measurement. The trigger logic required signals in Roman Pot scintillators on both sides of the STAR central barrel to ensure detection of scattered protons. In addition, at least 2 hits in the TOF detector were required, because in DPE the simplest state composed of charged particles is $\pi^{+}\pi^{-}$. Additionally the veto was exercised on signal in small BBC tiles and ZDCs to assure the rapidity gaps expected in DPE between final state protons and central diffractive system.

In order to select sample of CEP events the following selection criteria were used. First of all, proton tracks reconstructed on both sides of the STAR detector were required in analysis. Secondly, exactly two opposite-sign TPC tracks matched with hits in TOF were requested, both with $\arrowvert\eta\arrowvert<1$. From the difference in the time of detection of protons in Roman Pots the $z$-position of the vertex was reconstructed (resolution $\sigma_{z}=$~11~cm) and compared with the same quantity reconstructed in the TPC - maximum difference of $3\sigma_{z}$ was allowed. To reduce background originating from elastic proton-proton scattering overlapping with minimum bias pile-up events, it was checked whether proton tracks with near-beam momentum ($\xi<0.05$) are collinear $\big((\vec{p}_{1}+\vec{p}_{2})_{T} < 60~\textrm{MeV}/c\big)$ and, if yes, the event was dropped from further selection. To additionally clean selected sample a veto on minimum ionizing particle signal in large BBC tiles was imposed, which enlarged the rapidity gaps already required at the trigger level.

Identification of particles was performed using the energy loss of the tracks measured in the TPC. A pair of tracks was recognized as pions if d$E$/d$x$ of both was consistent at $3\sigma$ level with the mean energy loss of pion carrying the momentum equal to that of reconstructed track. In case of identification of kaon pair the selection was analogous with an additional requirement to have at least one track not consistent with pion hypothesis from d$E$/d$x$ at the $3\sigma$ level.

\begin{figure}[b!]
  \centering
  \begin{tabular}{c}~\\[-24pt]
    \includegraphics[width=0.41\linewidth]{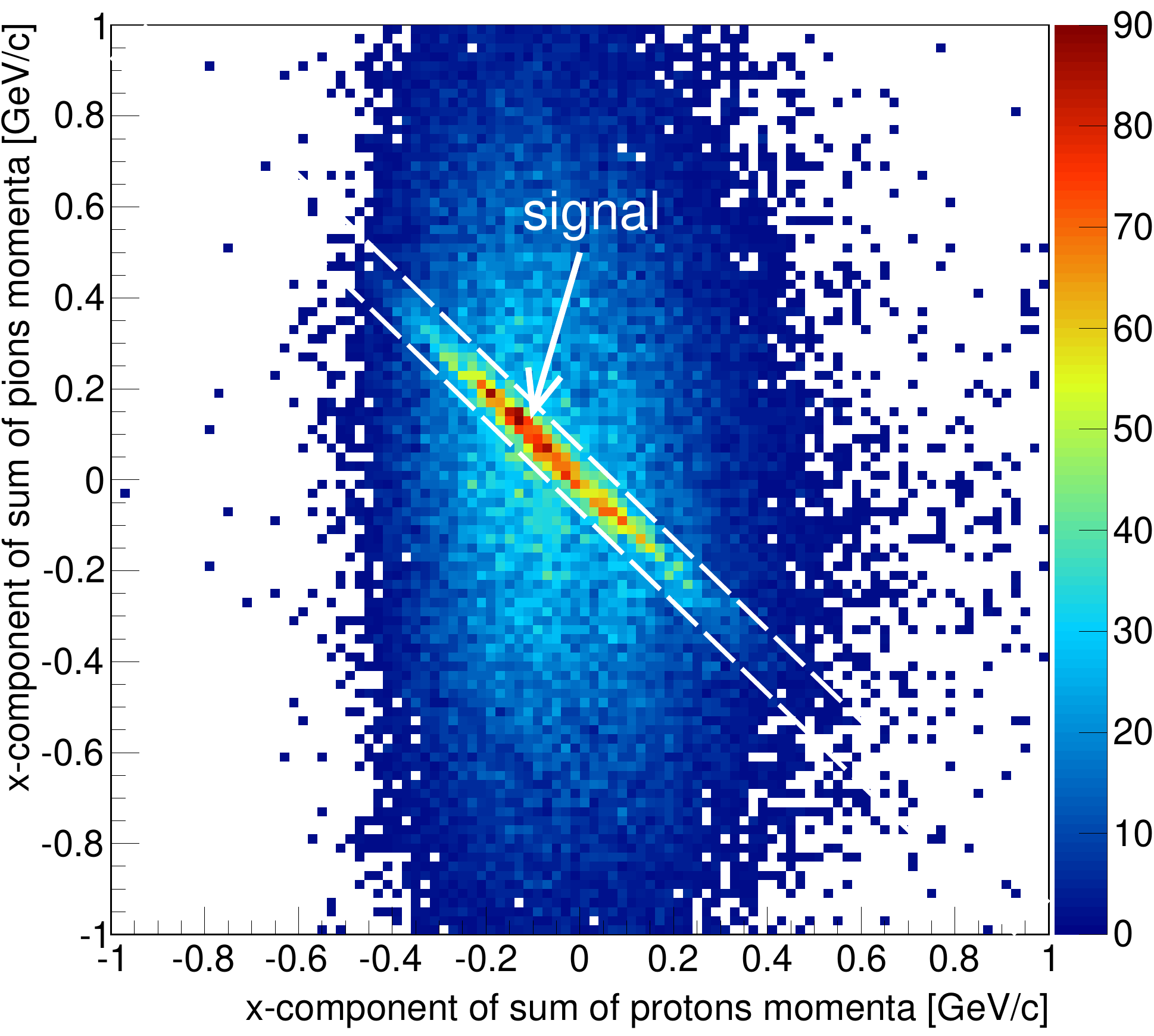}\label{Fig:correlation} \\[-5pt]
    \small (a)\\[-5pt]
  \end{tabular}
  \begin{tabular}{c}~\\[-29pt]
    \includegraphics[width=0.505\linewidth]{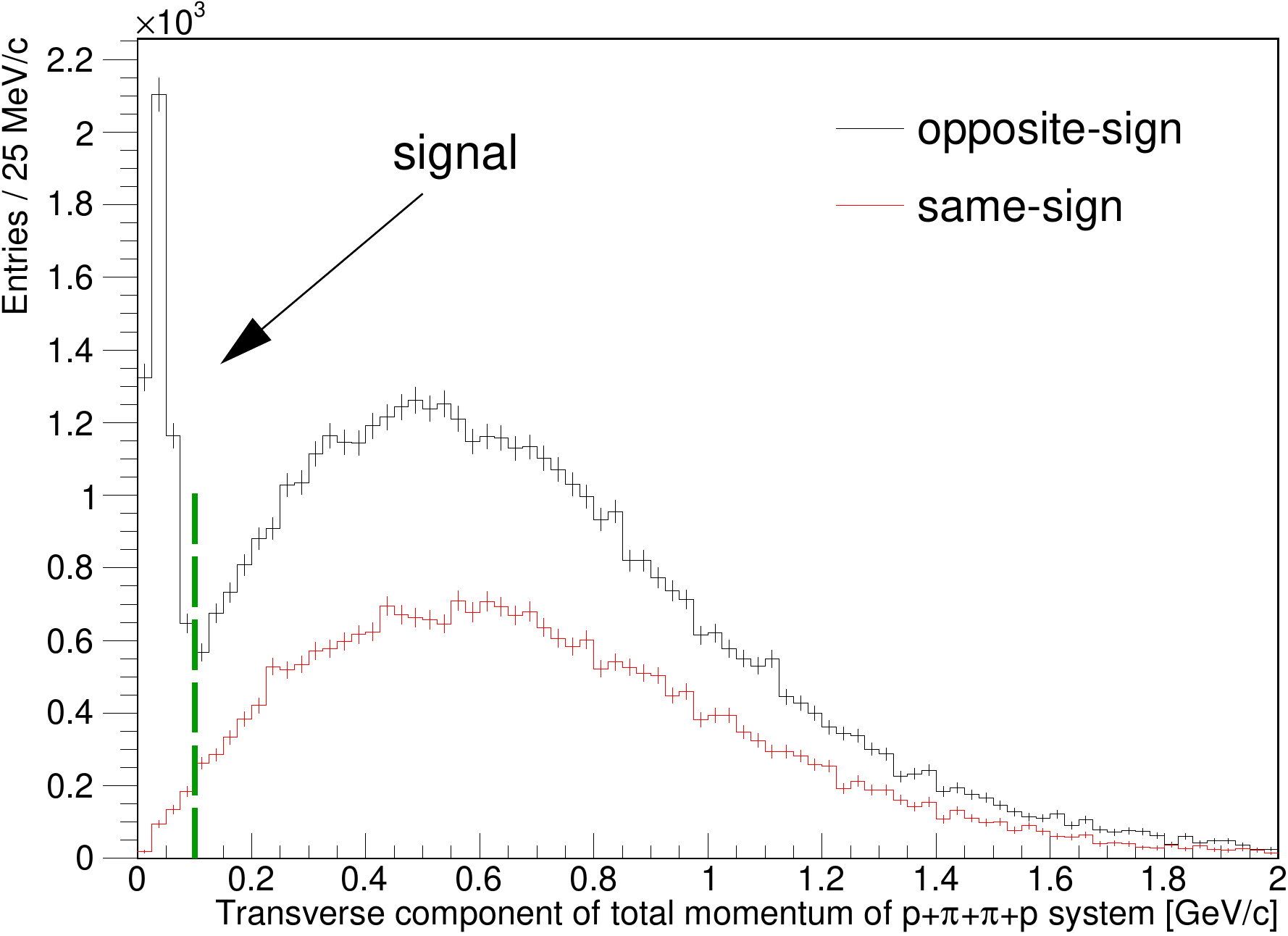}\label{Fig:missingPt} \\[-5pt]
    \small ~~~~~~~~~~~(b)\\[-5pt]
  \end{tabular}
  \caption{Correlation between $x$-component of the resultant momentum of opposite-sign pion tracks (vertical axis) and scattered protons (horizontal axis)~(a) and total transverse momentum of all final state particles (b).}\label{Fig:2}
\end{figure}

After event selection with the described set of cuts one can clearly see the signal events among remaining background on the correlation plot of $x$- and $y$-components of the sum of central tracks momenta and the sum of forward proton momenta, as presented in Figure~\ref{Fig:2}a. The anti-correlated signal region indicating events with balanced momentum distinctly stands out from the uncorrelated background. The momentum balance constraint is translated to the total transverse momentum ($p_{T}^{miss}$) cut, the final tool which helps to efficiently establish the exclusivity of measured diffractive system. Figure~\ref{Fig:2}b shows the distribution of $p_{T}^{miss}$ for pion pairs before the exclusivity cut, where the cut value of $100~\textrm{MeV}/c$ is marked with dashed green line. Prominent signal peak is visible for the opposite-sign pair events near the origin of the horizontal axis, whose offset from zero is caused by the angular divergence of the beam.

\begin{figure}
  \begin{tabular}{c}\hspace*{-0.03\linewidth}
    \includegraphics[width=0.5\linewidth]{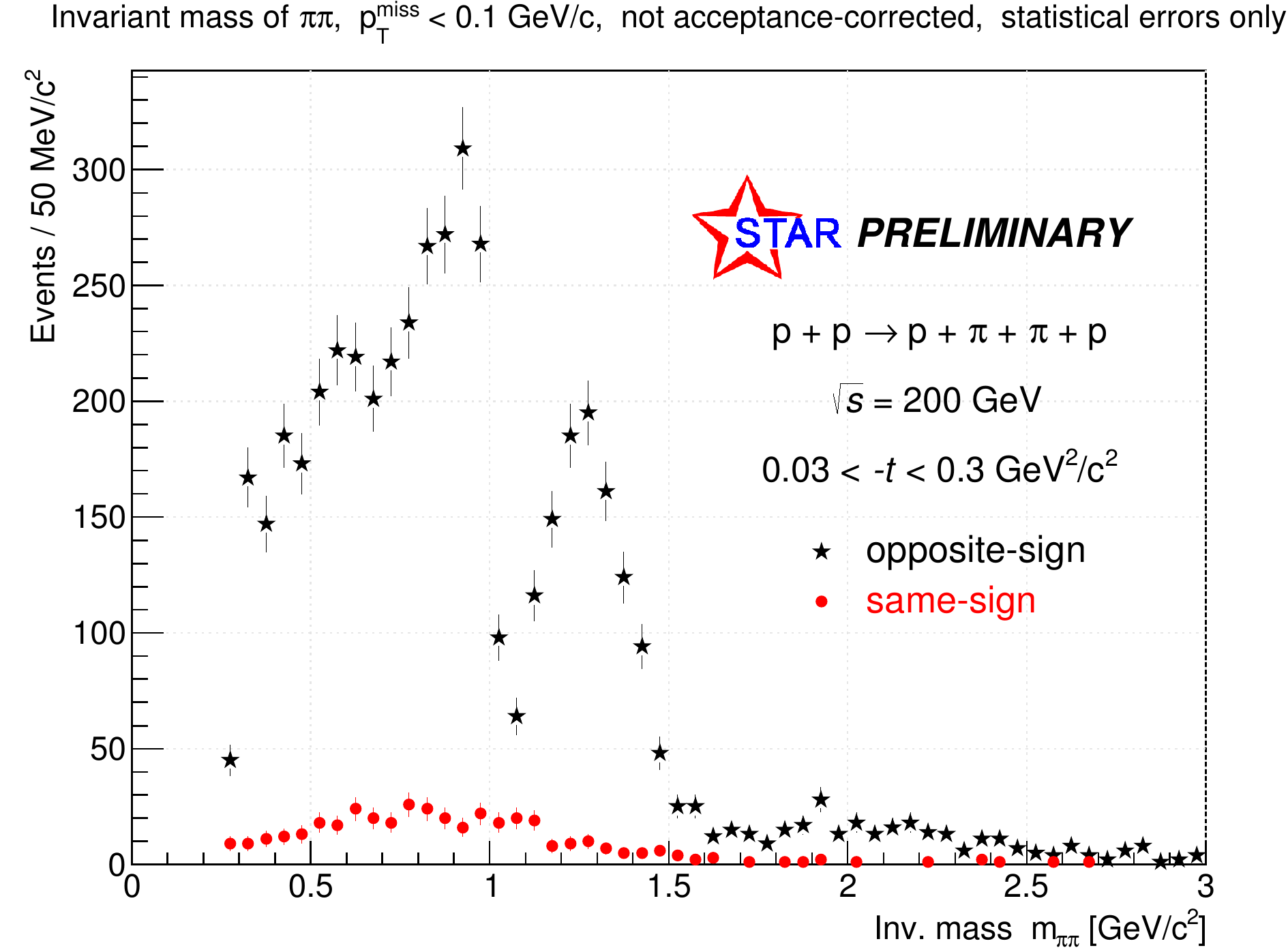}\label{Fig:pipi} \\[-10pt]
    \small (a)\\[-5pt]
  \end{tabular}\hspace{-0.02\linewidth}
  \begin{tabular}{c}%\hspace*{-0.05\linewidth}
    \includegraphics[width=0.5\linewidth]{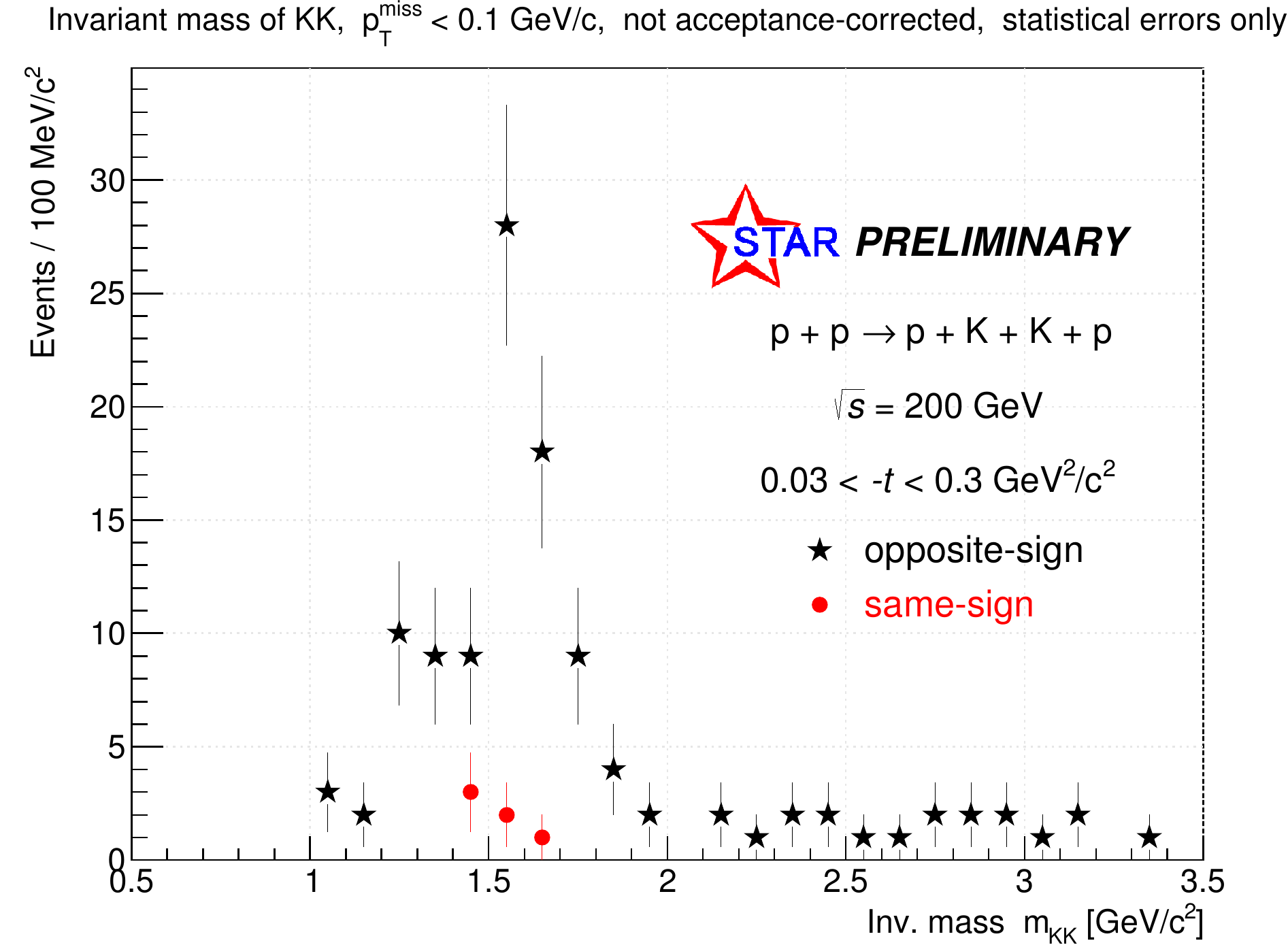}\label{Fig:KK}\\[-10pt]
    \small ~~~(b)\\[-5pt]
  \end{tabular}
  \caption{Invariant mass spectrum of exclusively produced pion~(a) and kaon~(b) pair. Distributions are not corrected for detector acceptance and efficiency. Error bars represent statistical uncertainty.}\label{Fig:mass}
\end{figure}

For all selected exclusive events, approximately $5\times10^{3}$ opposite-sign pion pairs and $10^{2}$ opposite-sign kaon pairs, the invariant mass of two TPC tracks was calculated. The results are shown in Figure~\ref{Fig:mass}.  In the invariant mass distribution of $\pi^{+}\pi^{-}$ (Figure~\ref{Fig:mass}a) one can observe broad structure extending from the production threshold to $1~\textrm{GeV}/c^{2}$, where rapid drop is found that might be attributed to the interference of $f_{0}(980)$ meson with the continuum. Between $1-1.5~\textrm{GeV}/c^{2}$ a~significant resonant structure is present which might represent the $f_{2}(1270)$ meson. This result is consistent with the previous measurement by STAR~\cite{Adamczyk} and agrees with predictions of some models~\cite{LebiedowiczSzczurek}. Lack of the peak of $\rho^{0}$ meson in $\pi^{+}\pi^{-}$ mass spectrum is a proof for insignificant contamination from Reggeon exchanges and photon-Pomeron interactions.

In the $K^{+}K^{-}$ channel the key feature is a prominent peak starting from $1.5~\textrm{GeV}/c^{2}$ which could be a $f_{0}(1500)$ state, a strong candidate for a mixture of glueball and the regular meson (quark-antiquark).

\section{SUMMARY AND OUTLOOK}

The STAR results on CEP of $\pi^{+}\pi^{-}$ and $K^{+}K^{-}$ has been presented and discussed. Analysis of the entire available statistics in the exclusive $\pi^{+}\pi^{-}$, $K^{+}K^{-}$, $p\bar{p}$ and $\pi^{+}\pi^{-}\pi^{+}\pi^{-}$ channels, involving final calibrations, optimized cuts and particle identification, is currently being finalized. With the number of nearly $2\times10^{5}$ exclusive $\pi^{+}\pi^{-}$ pairs, $2\times10^{3}$ exclusive $K^{+}K^{-}$ pairs and $10^{2}$ exclusive $p\bar{p}$ pairs the PWA in the first two channels is planned. Differential cross-sections with respect to various quantities will also be determined.

In run 2017 new data at higher center-of-mass energy of $\sqrt{s}=510$~GeV will be collected, which will allow comparison of the DPE in different kinematic regions. Also measurement of some other processes, such as exclusive jet production or $J/\psi$ photoproduction, will be enabled. The latter can give access to measurement of the generalized parton distributions in proton.

% Acknowledgement
\section{ACKNOWLEDGMENTS}
This work was partly supported by the National Science Centre of Poland under contract No. UMO-2015/18/M/ST2/00162. Author is supported by the Smoluchowski scholarship from the KNOW funding.

% References

\nocite{*}
\bibliographystyle{aipnum-cp}%
\bibliography{references}%

\end{document}